\documentclass[apj]{emulateapj}

\usepackage{graphicx,times}
\usepackage{subfigure}
\newcommand{\be}{\begin{equation}}
\usepackage{threeparttable}
\usepackage{booktabs}
\newcommand{\ee}{\end{equation}}
\newcommand{\bea}{\begin{eqnarray}}
\newcommand{\eea}{\end{eqnarray}}

\usepackage{enumerate}
\usepackage{amsmath}
\usepackage{cases}
\usepackage{longtable}
\usepackage{hyperref}
\usepackage{epstopdf}
\usepackage{amsmath,bm}
\usepackage{amssymb}
\usepackage{natbib}
\usepackage{morefloats}
\usepackage{multirow}
\usepackage{array}
\usepackage{verbatim}
\usepackage{setspace}

\begin{document}

\title{On the synchrotron spectrum of GRB prompt emission}

\author{Siyao Xu\altaffilmark{1}, Yuan-Pei Yang \altaffilmark{2,3}, and Bing Zhang \altaffilmark{2,4,5}}

\altaffiltext{1}{Department of Astronomy, University of Wisconsin, 475 North Charter Street, Madison, WI 53706, USA; 
Hubble Fellow, sxu93@wisc.edu}
\altaffiltext{2}{Kavli Institute for Astronomy and Astrophysics, Peking University, Beijing 100871, China}
\altaffiltext{3}{KIAA-CAS Fellow, yypspore@gmail.com}
\altaffiltext{4}{Department of Astronomy, School of Physics, Peking University, Beijing 100871, China}
\altaffiltext{5}{Department of Physics and Astronomy, University of Nevada Las Vegas, NV 89154, USA; zhang@physics.unlv.edu}

\begin{abstract}
The prompt emission spectrum of Gamma-ray bursts (GRB) is characterized by a smoothly joint broken power
law spectrum known as the Band function. The typical low-energy photon index is $\sim -1$, which poses challenge to the
standard synchrotron radiation models. We investigate the electron energy spectrum as a result of the 
interplay among  adiabatic stochastic acceleration (ASA),  particle injection, and  synchrotron cooling. 
In the ASA-dominated low-energy range,
 ASA enables an efficient hardening of the injected energy spectrum to approach a spectral index $-1$. 
In the synchrotron cooling-dominated high-energy range,
the injected high-energy electrons undergo  fast synchrotron cooling and have a softer photon spectrum. 
With the energy range of the injected electrons broadly covering both the ASA- and synchrotron cooling-dominated ranges, 
the resulting photon number spectrum has the low- and high-energy indices as 
$\alpha_s \sim -1$ and $\beta_s \sim -p/2-1$, respectively. 
The break energy is of the order of $\sim 100 $ keV, depending on the turbulence properties. 

\end{abstract}

\keywords{acceleration of particles - gamma-ray burst: general - radiation mechanisms: non-thermal - turbulence}

\section{Introduction}

After decades of observations, the radiation mechanism of gamma-ray bursts (GRBs) is still subject to debate
\citep[e.g.][]{Zh14,KuZ15,Pe15}. Even though it is agreed that some GRBs with narrow thermal-like spectrum
(e.g. GRB 090902B) is of a Compotonized photospheric origin (e.g. \citealt{Ryde10,Peer12}), the origin of the
so-called Band function component \citep{Band93,Pre00} commonly observed in GRBs is still unclear.
This is because the typical value of the low-energy photon index $\alpha \sim -1$ is not straightforwardly 
expected from either the thermal model \citep{Bel10,Deng14} or the synchrotron model \citep{Ree92,Kat94,Tav96}. 
This motivates the exploration of alternative solutions to the problem 
\citep[e.g.][]{Bra94,Lia97,Pree98,Me00,Med00,Ghi00,PeZ06, Daigne11, UZ14, Ben14, Asano15, Asa16}.

Electron acceleration is the fundamental physical ingredient 
for understanding the photon spectrum of the prompt emission. 
Turbulence is naturally expected in the magnetized GRB outflow 
via the magnetic reconnection.
The observational evidence for the turbulent GRB outflow can be found in,  
e.g., \citet{Bel98}.
The reconnection of regular magnetic fields can generate turbulence, which in turn facilitates rapid reconnection on 
macroscopic scales and determines the reconnection rate
\citep{LV99,Kow17}. 
Collision-induced turbulent reconnection in a moderately high-$\sigma$ regimes 
provides an efficient dissipation mechanism of the magnetic energy in the GRB outflow
\citep{Zh11,Deng15}.
Advances in the theory of particle acceleration in magnetohydrodynamic (MHD) turbulence 
provides us new insights into the GRB emission.

As a new stochastic acceleration mechanism identified by 
\citet{BruL16}, 
electrons entrained on turbulent magnetic field lines, i.e., 
with their Larmor radii smaller than the characteristic length scale of magnetic fields,
are randomly accelerated and decelerated. 
It is different from the gyroresonant interaction with MHD waves, 
because 
(1) it originates from the second adiabatic invariant; 
(2) it accelerates non-resonant particles; 
(3) the particles undergo the first-order Fermi process and thus
a large fractional energy change within individual turbulent eddies.
\citet{XZg17}
(hereafter Paper \uppercase\expandafter{\romannumeral1}) 
first applied this adiabatic stochastic acceleration (ASA) to the GRB context, 
showing that it can efficiently harden any initial energy distribution of electrons at low energies, 
where acceleration dominates over synchrotron cooling.

The observed GRB spectra during the prompt phase are characterized by a smoothly connected broken power law
\citep{Band93,Pre00}.
While the hard low-energy spectrum can be accounted for by the ASA, as shown in 
Paper \uppercase\expandafter{\romannumeral1}, 
the soft spectrum at high energies, where the ASA is insignificant, should result from different physical processes. 
One important physical process that was not included in \cite{XZg17} was fast synchrotron cooling of 
instantaneously accelerated 
electrons, which is believed to play a significant role in GRB physics \citep[e.g.][]{Sar98,UZ14}.
Based on this knowledge and the findings in Paper \uppercase\expandafter{\romannumeral1}, 
in this paper, we introduce the particle injection, which not only serves as the particle source for the ASA
at low energies,  
but also contributes to developing the high-energy spectrum, together with synchrotron cooling. 
The injected particles can result from a separate instantaneous acceleration process. 
Therefore, apart from the continuous ASA and synchrotron cooling, we also incorporate the instantaneous acceleration 
through invoking particle injection. 
Our purpose is to model the interplay among different processes and investigate the physical origin of 
the synchrotron spectrum of GRB prompt emission.

This paper is organized as follows. In Section 2, we analyzed the time evolution of the electron energy spectrum under the effects 
of the ASA, injection, and radiation losses. 
In Section 3, we provide the synchrotron spectrum in both the slow- and fast-cooling regimes, 
as well as the characteristic spectral parameters. 
The discussion and summary of our results are in Section 4 and 5, respectively.

\section{Adiabatic stochastic acceleration and steady injection }\label{sec: ansta}

To study the time evolution of an electron energy spectrum $N(E,t)$,
we consider the kinetic equation of electrons 
\begin{equation}\label{eq: soasi}
   \frac{\partial N}{\partial t} =  a_2 \frac{\partial}{\partial E} \Big(E\frac{\partial (EN)}{\partial E}\Big)  
    + \beta \frac{\partial (E^2N)}{\partial E} + Q(E).
\end{equation}
The first term on the right-hand side accounts for 
the ASA introduced in Paper \uppercase\expandafter{\romannumeral1}. 
It arises in the trans-Alfv\'{e}nic turbulence 
\citep{GS95}
with balanced magnetic and kinetic energies. 
The acceleration is stochastic  
as the particles encounter the statistically equally distributed
reconnecting and dynamo regions in MHD turbulence. 
Resulting from the second adiabatic invariant, they are accelerated when the magnetic field lines shrink during 
the turbulent reconnection, 
and decelerated when the field lines are stretched via the turbulent dynamo. 
On the other hand, the acceleration/deceleration within individual turbulent eddies 
is the first-order Fermi process, 
in an analogy that the particles are 
trapped between approaching/receding mirrors.

The acceleration rate $a_2$ is expressed as (Paper \uppercase\expandafter{\romannumeral1})
\begin{equation}\label{eq: a2}
    a_2 = \xi \frac{u_\text{tur}}{l_\text{tur}},
\end{equation}
where $l_\text{tur}$ and $u_\text{tur}$ are the injection scale and velocity of the trans-Alfv\'{e}nic turbulence,
$\xi = \Delta E / E$ is the fractional energy change within each turbulent eddy-turnover time 
$\tau_\text{tur} = l_\text{tur}/u_\text{tur}$.
In non-relativistic turbulence, $\tau_\text{tur}$ is sufficiently long  
for particles with their Larmor radii $r_L$ smaller than $l_\text{tur}$ to have many crossings within individual eddies.
The significant energy change during every interaction time $\tau_\text{tur}$
leads to a large energy diffusion rate 
\begin{equation}
    D_{\rm EE} = \frac{\xi  E^2 }{\tau_\text{tur}}, 
\end{equation}
and thus efficient acceleration, 
which is in contrast to the pitch-angle scattering of particles by MHD waves with a small energy change per collision. 
In the case of highly relativistic turbulence, 
one crossing in a turbulent eddy is efficient enough to drastically change the particle energy with 
$\xi$ of the order $\gamma_\text{tur}^2$.
It comes from twice 
Lorentz transformations, where $\gamma_\text{tur}$ is the turbulence Lorentz factor.

The second term on the right-hand side of Eq. \eqref{eq: soasi} describes the 
synchrotron and synchrotron-self-Compton (SSC) losses. 
The parameter $\beta$ is defined as
\begin{equation}\label{eq: betp}
\begin{aligned}
    \beta &= \frac{P_\text{syn} + P_\text{ssc}}{E^2} = \frac{P_\text{syn} (1+Y)}{E^2}   \\
             &= \frac{\frac{4}{3} \sigma_T c U_B \gamma_e^2 (1+Y)}{(\gamma_e m_e c^2)^2}
               = \frac{\sigma_T c B^2 (1+Y)}{ 6 \pi (m_e c^2)^2},
\end{aligned}
\end{equation}
where $P_\text{syn}$ and $P_\text{ssc}$ are the powers of the synchrotron and SSC radiation, respectively, 
with $Y$ as the ratio between $P_\text{ssc}$ and $P_\text{syn}$.
Other parameters are the magnetic field strength $B$, 
the magnetic energy density $U_B = B^2 / (8 \pi)$, 
the Thomson cross section $\sigma_T$, the electron Lorentz factor $\gamma_e$, the electron rest mass
$m_e$, and the speed of light $c$.

The third term in Eq. \eqref{eq: soasi} represents the 
steady injection of a constant electron spectrum $Q(E) = C E^{-p}$ with a power-law index $p$. 
At some fixed reference energy, the constant parameter $C$ corresponds to 
the number of the injected electrons per unit energy per unit time.
The steady injection of new particles was not included in Paper \uppercase\expandafter{\romannumeral1}. 
By taking it into account, we involve 
other possible instantaneous acceleration mechanisms 
that produce power-law electron spectra.
The instantaneous acceleration processes, taking place in a separate region, 
provide the particle source in the magnetized and turbulent medium 
for further ASA as described by the first term in Eq. \eqref{eq: soasi}.
The remaining terms in the general kinetic equation 
(see e.g. \citealt{Gin57,Kar62,Che78})
are excluded as they are of marginal importance to our analysis.

In Appendix \ref{app}, we present a more general analysis for the evolving electron energy distribution under the 
effects of both stochastic and systematic acceleration. 
It recovers the situation considered here when the stochastic acceleration is the dominant acceleration process 
in the turbulence region.

By substituting $EN =  \exp(-\epsilon E) u(x, \tau)$, $\epsilon = \beta / a_2$, $x=\ln E$, and $\tau = a_2 t$
into Eq. \eqref{eq: soasi}, we have 
\begin{equation}\label{eq: simeqe}
      \frac{\partial u}{\partial \tau}   =    \frac{\partial^2 u}{\partial x^2} -   \frac{E}{E_\text{cf}} \frac{\partial u}{\partial x}  
      + \frac{C}{a_2} \exp{(\epsilon E)} E^{-p +1},
\end{equation}
where we define the cutoff energy for the ASA as 
\begin{equation}\label{eq: coeg}
      E_\text{cf} = \frac{a_2}{\beta}  = \frac{1}{\epsilon},
\end{equation}
corresponding to the balance between the ASA and the energy losses due to the radiation.
Under the condition $E \ll E_\text{cf}$, Eq. \eqref{eq: simeqe} approximately becomes 
\begin{equation}\label{eq: inhoh}
    \frac{\partial u}{\partial \tau}   = \frac{\partial^2 u}{\partial x^2}  +  C^\prime \exp{\big((-p +1)x\big)}, 
\end{equation}
where $C^\prime = C/a_2$. 
Eq. \eqref{eq: inhoh} has the same form as an inhomogeneous heat equation.
With the initial condition $u(x,0)=0$, its solution is given by 
(see \citealt{Ca84}),
\begin{equation}\label{eq: gen}
\begin{aligned}
    u(x,\tau) = & C^\prime \int_0^\tau  \frac{1}{2\sqrt{\pi (\tau-s)}}    \\
                     & \int_{y_l}^{y_u} \exp\bigg(-\frac{(x-y)^2}{4(\tau-s)}\bigg)  \exp{\big((-p +1)y\big)} dy ds,
\end{aligned}
\end{equation}
where $y_l = \ln E_l$ and $y_u = \ln E_u$, with  
$E_l$ and $E_u$ as the lower and upper limits of the injected energy spectrum.
We note that when $s=0$, 
the above solution becomes  
\begin{equation}\label{eq: spe}
\begin{aligned}
  &  u^*(x,\tau) = u(x,\tau)_{s=0} =     \frac{C^\prime}{2\sqrt{\pi \tau}}    \\
                     & \int_{y_l}^{y_u} \exp\bigg(-\frac{(x-y)^2}{4\tau}\bigg)  \exp{\big((-p +1)y\big)} dy .
\end{aligned}
\end{equation}
The corresponding electron energy spectrum is 
\begin{equation}\label{eq: p1ns}
\begin{aligned}
 &  N^*(E,\tau)  =  E^{-1}  \frac{C^\prime}{2 \sqrt{\pi \tau}}  \exp\Big(-\frac{E}{E_\text{cf}}\Big)  \\
 &   \int_{y_l}^{y_u}   \exp\bigg(-\frac{(\ln E - y)^2}{4 \tau}\bigg) \exp \big((-p+1)y\big) dy . 
\end{aligned}
\end{equation}
As discussed in Paper \uppercase\expandafter{\romannumeral1},
this is the case of impulsive injection with the initial condition  
\begin{equation}\label{eq: inpw}
   N^*(E,0) = C^\prime E^{-p} \exp \Big(-\frac{E}{E_\text{cf}}\Big), ~~ E_l < E< E_u.
\end{equation}

As regards the steady injection considered here, Eq. \eqref{eq: gen} can be treated as the superposition 
of a series of $u^*(x,\kappa)$ (Eq. \eqref{eq: spe}) with $\kappa$ ranging from $0$ to $\tau$, i.e., 
\begin{equation}\label{eq: uint}
    u(x,\tau) = \int_0^\tau u^*(x,\kappa) d\kappa.
\end{equation}
When $\tau$ is small, 
by integrating the simplified form of $u^*(x,\kappa)$ at the limit of small $\kappa$
\begin{equation}
    u^*(x,\kappa)_\text{small $\kappa$} = \frac{C^\prime}{2\sqrt{\pi \kappa}}  \exp \big((-p+1)x\big), 
\end{equation}
Eq. \eqref{eq: uint} yields: 
\begin{equation}
    u(x,\tau)  = 
    \frac{C^\prime \sqrt{\tau}}{\sqrt{\pi}}  \exp \big((-p+1)x\big). 
\end{equation}
Then we obtain the short-time asymptotic energy spectrum: 
\begin{equation}\label{eq: nsht}
   N(E,\tau)  =  \frac{C^\prime \sqrt{\tau}}{\sqrt{\pi}} E^{-p} \exp \Big(-\frac{E}{E_\text{cf}}\Big).
\end{equation}
The spectral form is governed by the injected particle distribution. 
Notice that 
different from the situation with an impulsive source
($N^*(E,\tau)$ in Eq. \eqref{eq: p1ns}), 
the above expression shows $N(\tau) \propto \sqrt{\tau}$.
It means 
that the steady energy injection leads to an increase of the electron number per unit energy with time.

When $\tau$ is sufficiently large, we integrate the reduced form of 
$u^*(x,\kappa)$ (Eq. \eqref{eq: spe}) at the limit of a large $\kappa$, which is  
\begin{equation}
\begin{aligned}
    u^*(x,\kappa)_\text{large $\kappa$} = & \frac{C^\prime }{2(-p+1)\sqrt{\pi \kappa}}    \\
    & \Big[\exp\big((-p+1)y_u\big) - \exp\big((-p+1)y_l\big)\Big], 
\end{aligned}
\end{equation}
and find (Eq. \eqref{eq: uint})
\begin{equation}
\begin{aligned}
    u(x,\tau)  
   = & \frac{C^\prime \sqrt{\tau} }{(-p+1)\sqrt{\pi }}    \\
      & \Big[\exp\big((-p+1)y_u\big) - \exp\big((-p+1)y_l\big)\Big]. 
\end{aligned}     
\end{equation}
So at a long time,
the energy spectrum asymptotically approaches
\begin{equation}\label{eq: fism}
     N(E,\tau)   
     =  \frac{C^\prime (E_u^{-p+1} - E_l^{-p+1}) \sqrt{\tau} }{(-p+1)\sqrt{\pi }}    
         E^{-1} \exp \Big(-\frac{E}{E_\text{cf}}\Big)  .
\end{equation}
The ASA plays a dominant role in regulating the spectral form 
and the resulting spectral index converges to a universal value of $-1$ 
at $E<E_\text{cf}$, irrespective of the injected power-law index $p$. 
This is valid for both impulsive and steady injection, 
but obviously in the latter situation it requires relatively longer time for the overall spectrum to 
comply with the universal power-law.

\begin{figure}[htbp]
\centering
\includegraphics[width=9cm]{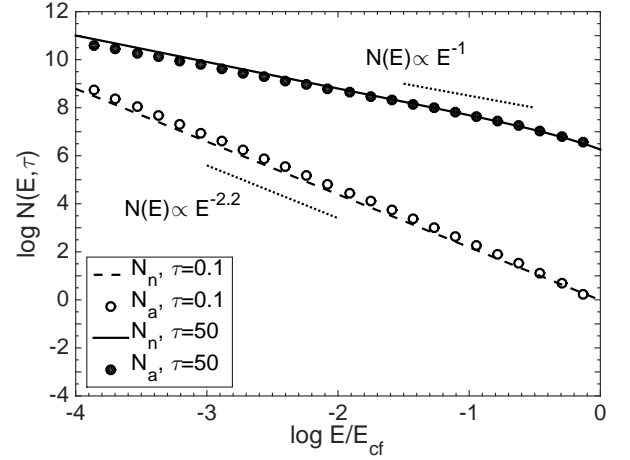}
\caption{ The electron energy spectrum at different times. $N_n$ denotes the numerical results by solving Eq. \eqref{eq: simeqe}.
$N_a$ represents the analytical approximations, for which we use Eq. \eqref{eq: nsht} at $\tau=0.1$ and Eq. \eqref{eq: fism} at $\tau= 50$. }
\label{fig: num}
\end{figure}

We present the above analytically derived approximate results in Fig. \ref{fig: num}, 
in a good agreement with the numerical solutions to Eq. \eqref{eq: simeqe} at $E< E_\text{cf}$. 
For the numerical illustration, 
we adopt a steep spectral slope $p = 2.2$ of the injected electrons and $C^\prime =10$ in arbitrary units.
The value of $C^\prime$ does not affect the spectral form, but only the amplitude of the spectrum. 
$E$ is normalized by $E_\text{cf}$, and $N(E,\tau)$ is in arbitrary units.
The numerical results indeed confirm that at an early time, the form of $N(E,\tau)$ 
is dictated by the injected spectrum. 
At a later time, i.e., several tens of $1/a_2$,
a hard spectrum with the index approaching $-1$ is developed below $E_\text{cf}$
under the influence of the ASA.

\section{Distribution of high-energy electrons and the synchrotron spectrum}
\subsection{Slow- and fast-cooling regimes }

The electrons injected into the trans-Alfv\'{e}nic turbulence undergo the ASA, 
and consequently their original power-law distribution evolves to a hard spectrum.  
The hard spectrum can only extend over the energy range below $E_\text{cf}$, 
since above $E_\text{cf}$, the ASA becomes ineffective due to the more important 
radiation losses of electrons. 
Therefore, in the following analysis on the high-energy portion of the electron spectrum, 
we assume that it is only shaped by the particle injection and synchrotron losses ($Y=0$).
The narrow relativistic Maxwellian distribution which can be piled up above $E_\text{cf}$ 
\citep{Sch84,Sch85}
is neglected.

The high-energy electron spectrum 
falls in the cooling regimes as discussed in 
\citet{Sar98}. 
To evaluate the synchrotron cooling effect, we define a critical energy $E_c$,
\begin{equation}
     E_c = P_\text{syn} t.
\end{equation}
Combining it with Eq. \eqref{eq: betp}, there is 
\begin{equation}\label{eq: enecol}
   E_c = \frac{1}{\beta t},
\end{equation}
which decreases with time $t$. 
The electrons with $E>E_c>E_\text{cf}$ cool down to $E_c$ within the time $t$. 

Depending on the relation between $E_c$ and $E_l$, which is the 
lower limit of the energies of the injected electrons, 
we next separately discuss the slow-cooling regime with $E_c > E_l$ and the fast-cooling regime with $E_c < E_l$.

(1) {\it Slow cooling } 

The energy spectrum of the electrons with $E>E_\text{cf}$ follows the kinetic equation 
\begin{equation}
     \frac{\partial N}{\partial t} =  \beta \frac{\partial (E^2N)}{\partial E} +  C E^{-p} , 
\end{equation}
where the acceleration term is absent. 
With the initial condition $N(E,0)=0$, its solution was provided by 
\citet{Kar62},
\begin{equation}
    N(E,t) = \frac{C E^{-(p+1)}}{\beta(p-1)} [ 1 -   (1- \beta t E)^{p-1} ],
\end{equation}
which is asymptotically a broken power-law, 
\begin{subnumcases}
     {  N(E,t) \approx \label{eq: slcin}}
     C t E^{-p}, ~~~~~~~~~~    E_\text{cf} < E \ll E_c, \\
     \frac{CE^{-(p+1)}}{\beta (p-1)}, ~~~~~~~~ E \gg E_c.  \label{eq: cols}
\end{subnumcases}
This spectrum can only arise at high energies above $E_\text{cf}$ with negligible acceleration effect.

\begin{figure*}[htbp]
\centering
\subfigure[Case (\romannumeral1)]{
   \includegraphics[width=8.5cm]{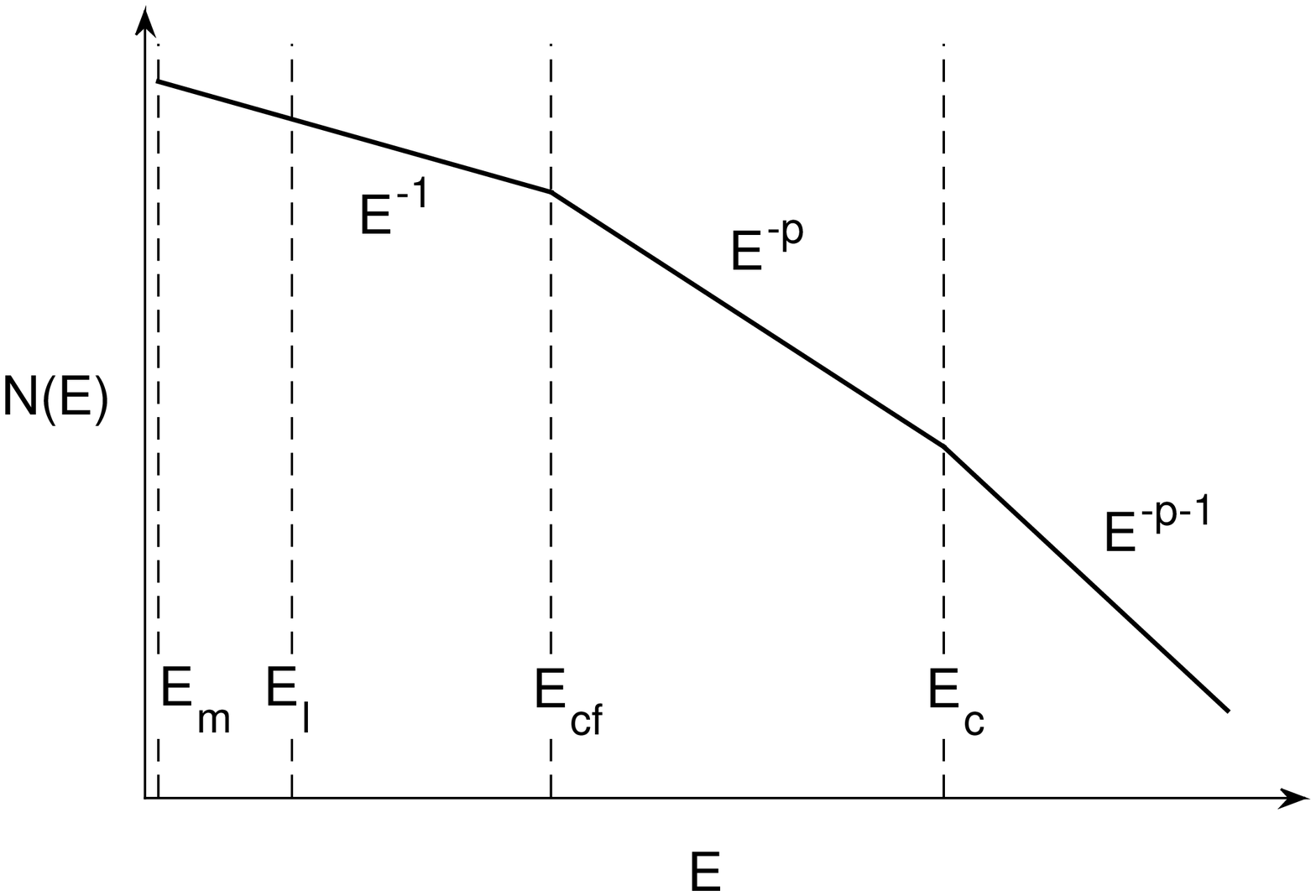}\label{fig: sloc1}}
\subfigure[Case (\romannumeral1)]{
   \includegraphics[width=8.5cm]{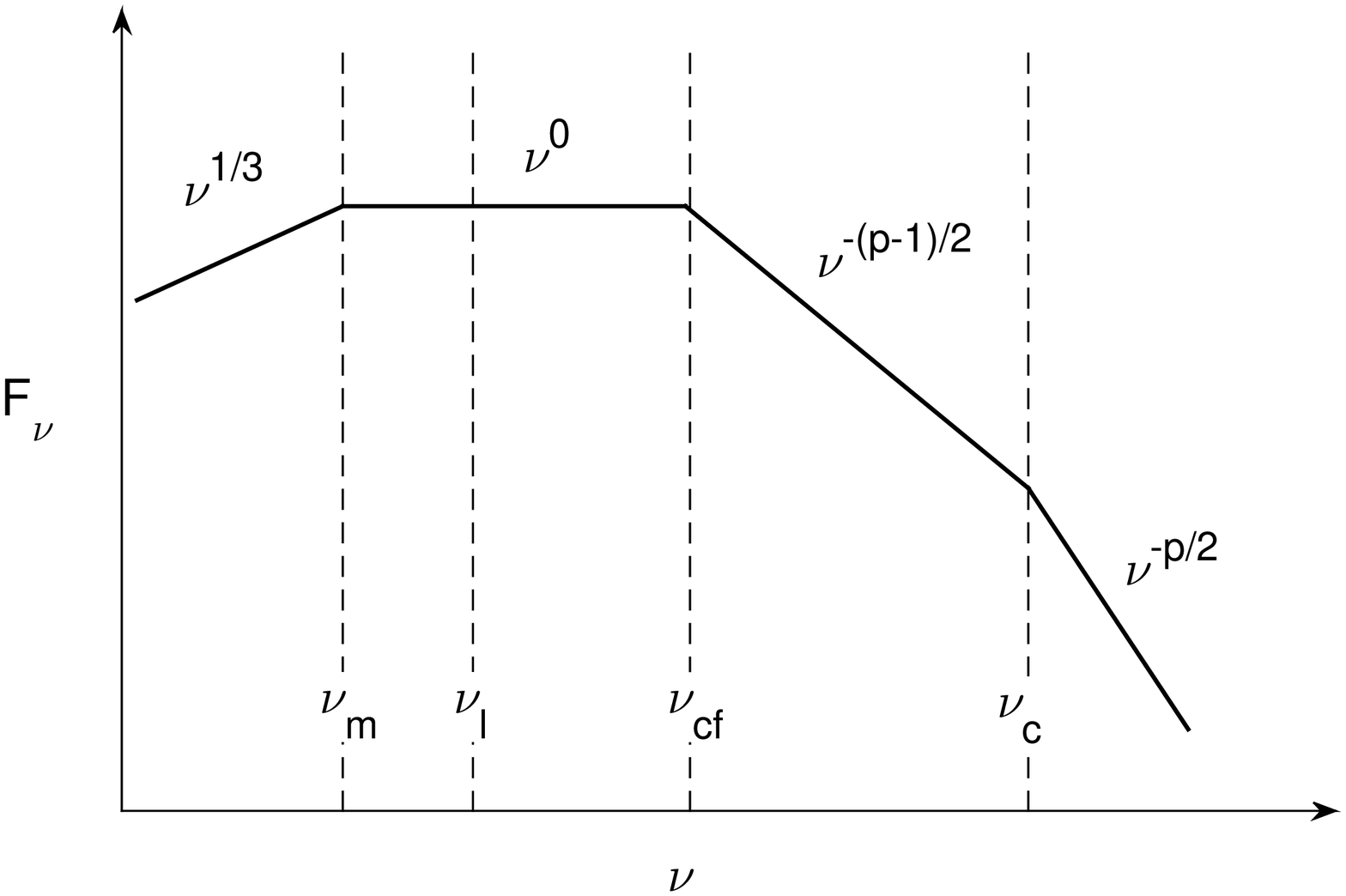}\label{fig: sloc1s}}   
\subfigure[Case (\romannumeral2)]{
   \includegraphics[width=8.5cm]{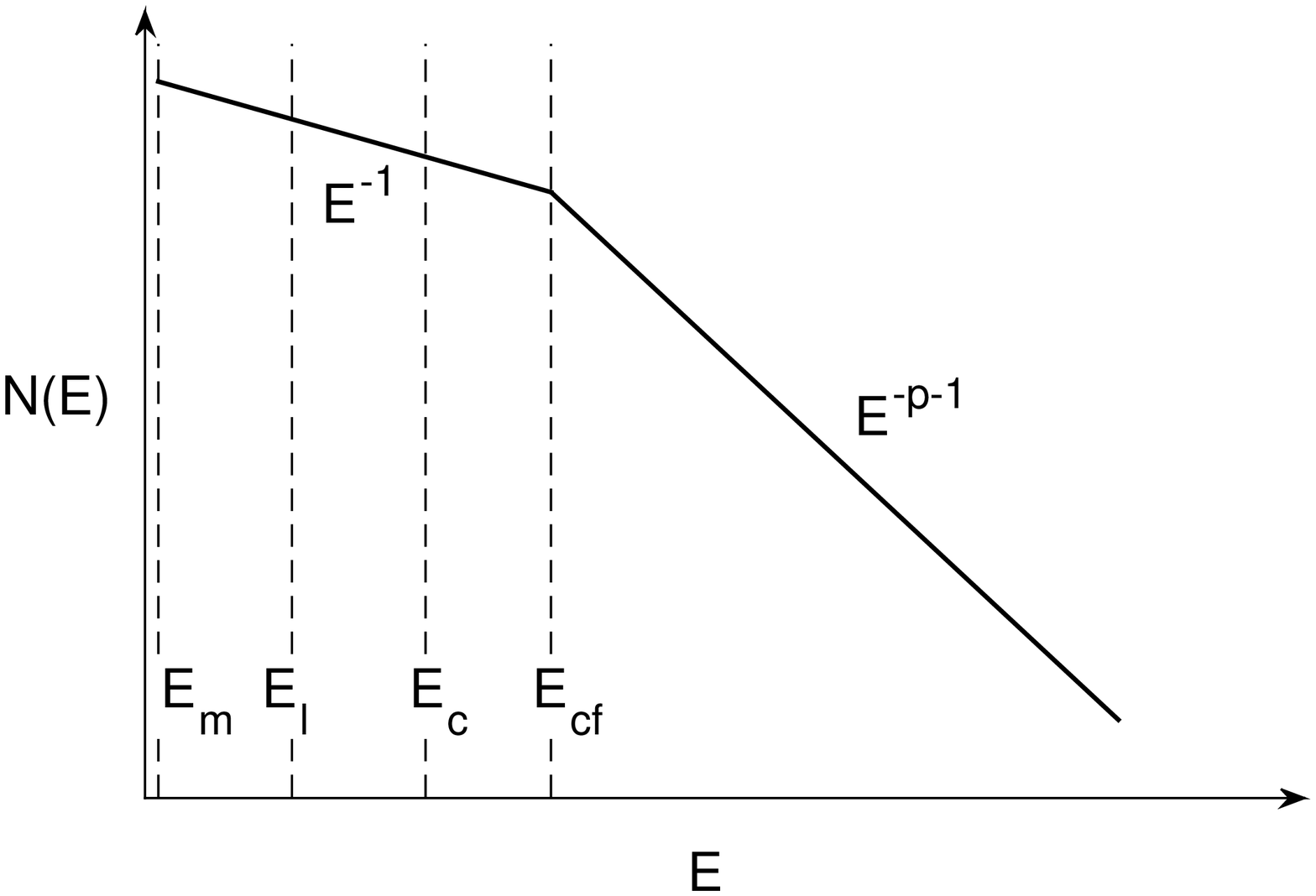}\label{fig: sloc2}}
\subfigure[Case (\romannumeral2)]{
   \includegraphics[width=8.5cm]{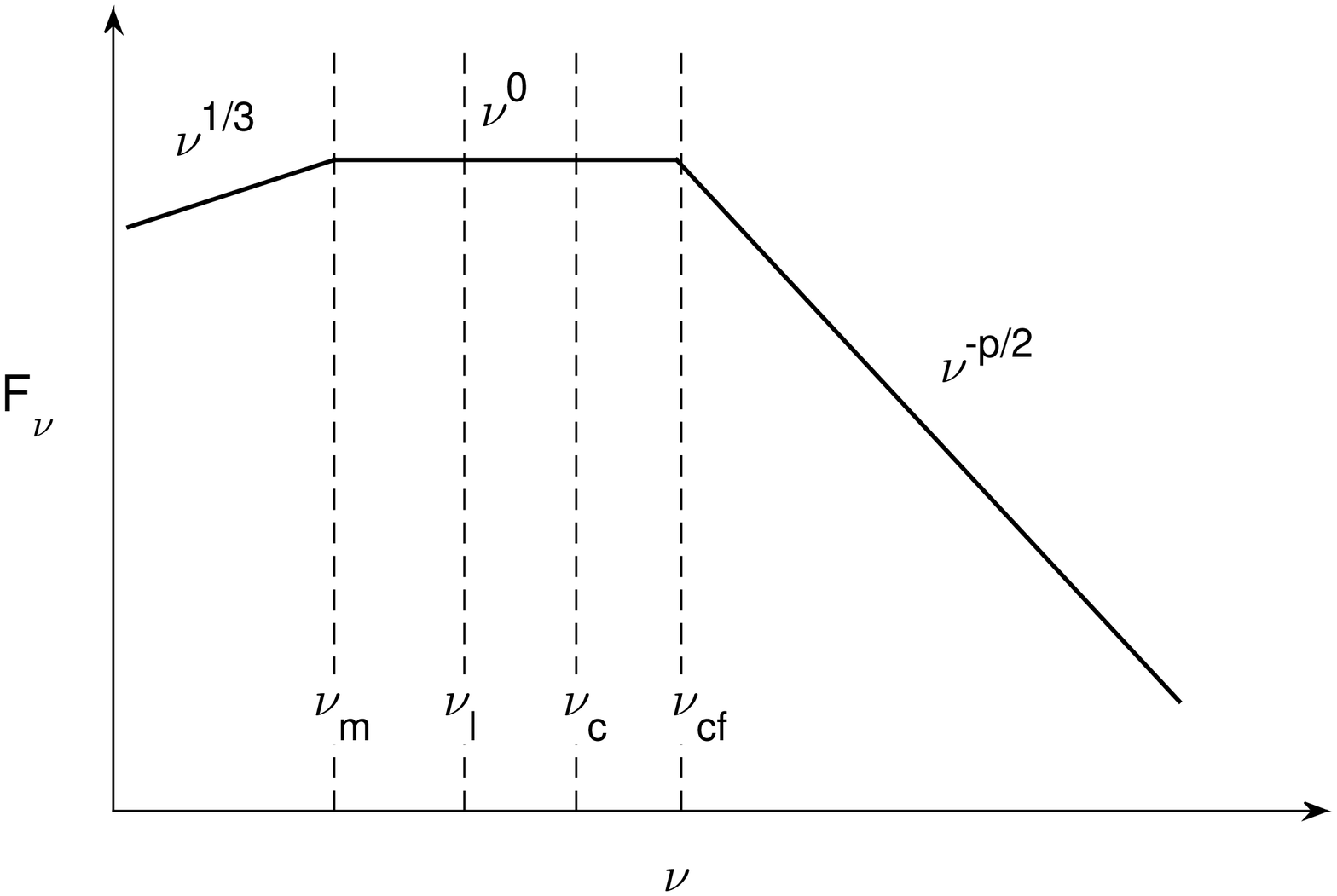}\label{fig: sloc2s}} 
\caption{ The electron energy spectrum ((a) and (c)) and the corresponding synchrotron spectrum ((b) and (d)) in the slow-cooling regime. }
\label{fig: slow}
\end{figure*}

Depending on the time $t$, 
the relation between $E_\text{cf}$ and $E_c$ varies. Correspondingly, 
there exist two scenarios for the electron energy spectrum in the slow-cooling regime, 
as shown in Fig. \ref{fig: sloc1} and \ref{fig: sloc2}, respectively.

At a early time with 
$E_c > E_\text{cf}$ (Case (\romannumeral1), see Fig. \ref{fig: sloc1}), i.e., $t< 1/a_2$, 
the time is so insufficient that only the very high-energy electrons undergo significant synchrotron cooling 
and have their distribution complying with Eq. \eqref{eq: cols}. 
With the decrease of $E_c$ with time, the spectral steepening extends to lower energies. 
But as soon as $t > 1/a_2$, i.e., $E_c < E_\text{cf}$, is reached
(Case (\romannumeral2), see Fig. \ref{fig: sloc2}), 
the ASA dominates over the synchrotron cooling, and no further steepening is expected below $E_\text{cf}$. 
Besides, 
as a result of the energy diffusion facilitated by the ASA, 
the minimum energy $E_m$ is 
\begin{equation}\label{eq: eml}
    E_m =  E_l \exp (-2\sqrt{\tau}), 
\end{equation}
which is smaller than the lower bound $E_l$ to the 
energy range of the injected electrons.
Therefore, the injected electron spectrum is extended over a broader energy range.

Given the electron energy spectrum 
\begin{equation}
    N(E) \propto E^{-p}  ,
\end{equation}
the synchrotron photon number spectrum in terms of frequency $\nu$ is 
\citep{Ry79}
\begin{equation}
    N(\nu) \propto \nu^\alpha , ~~
   \alpha =  -\frac{p+1}{2} .
\end{equation}
The flux $F_\nu \propto \nu N(\nu)$ corresponding to the above two cases are: 
\\
Case (\romannumeral1) $E_l < E_\text{cf} < E_c$,
\begin{subnumcases}
 {F_\nu=}
F_{\nu,\text{max}} \Big(\frac{\nu}{\nu_m}\Big)^\frac{1}{3}  ,~~~~~~~~~~~~~~~~~~~~~~~~\nu<\nu_m,\\
F_{\nu,\text{max}},~~~~~~~~~~~~~~~~~~~~~~~~~~~~\nu_m < \nu <\nu_\text{cf}, \\
F_{\nu,\text{max}} \Big(\frac{\nu}{\nu_\text{cf}}\Big)^{-\frac{p-1}{2}} ,~~~~~~~~~~\nu_\text{cf} < \nu < \nu_c, \\
F_{\nu,\text{max}} \Big(\frac{\nu_c}{\nu_\text{cf}}\Big)^{-\frac{p-1}{2}} \Big(\frac{\nu}{\nu_c}\Big)^{-\frac{p}{2}} ,~~~~~~\nu_c < \nu.
\end{subnumcases}
Case (\romannumeral2) $E_l < E_c < E_\text{cf}$,
\begin{subnumcases}
 {F_\nu=}
F_{\nu,\text{max}} \Big(\frac{\nu}{\nu_m}\Big)^\frac{1}{3}  ,~~~~~~~~~~~~~~~~~~~~~~~~\nu<\nu_m,\\
F_{\nu,\text{max}},~~~~~~~~~~~~~~~~~~~~~~~~~~~~\nu_m < \nu <\nu_\text{cf}, \\
F_{\nu,\text{max}} \Big(\frac{\nu}{\nu_\text{cf}}\Big)^{-\frac{p}{2}}, ~~~~~~~~~~~~~~\nu_\text{cf} < \nu < \nu_c.
\end{subnumcases}
They are illustrated in 
Fig. \ref{fig: sloc1s} and \ref{fig: sloc2s}.
The characteristic frequencies
$\nu_m$, $\nu_l$, $\nu_\text{cf}$, and $\nu_c$ are related to  
$E_m$, $E_l$, $E_\text{cf}$, and $E_c$.
In the presence of the flat segment of $F_\nu$, 
we use $F_{\nu,\text{max}}$ to represent the constant peak flux over the frequency range $(\nu_m, \nu_\text{cf})$. 
As a characteristic feature of the synchrotron radiation, 
the low-frequency tail $F_\nu \propto \nu^{1/3}$ is determined by the synchrotron single-particle emission spectrum 
and is independent of the underlying electron distribution
\citep{MR93,Kat94}.

Regarding the $\nu F_\nu$ spectrum, 
the spectrum in Case (\romannumeral1) peaks at $\nu_c$
under the condition $2<p<3$. 
The spectral peak in Case (\romannumeral2) is at $\nu_\text{cf}$ as long as $p>2$.
It implies that the spectral peak of $\nu F_\nu$ shifts from large to small frequencies with time, 
until settling at $\nu_\text{cf}$.

(2) {\it Fast cooling} 

In fact, only the electrons with the energies exceeding $E_\text{cf}$ are in the fast-cooling regime. 
Depending on the relations between $E_\text{cf}$ and $E_l$, 
there are also different cases for the electron energy spectrum.

\begin{figure*}[htbp]
\centering   
\subfigure[Case (\romannumeral1)]{
   \includegraphics[width=8.5cm]{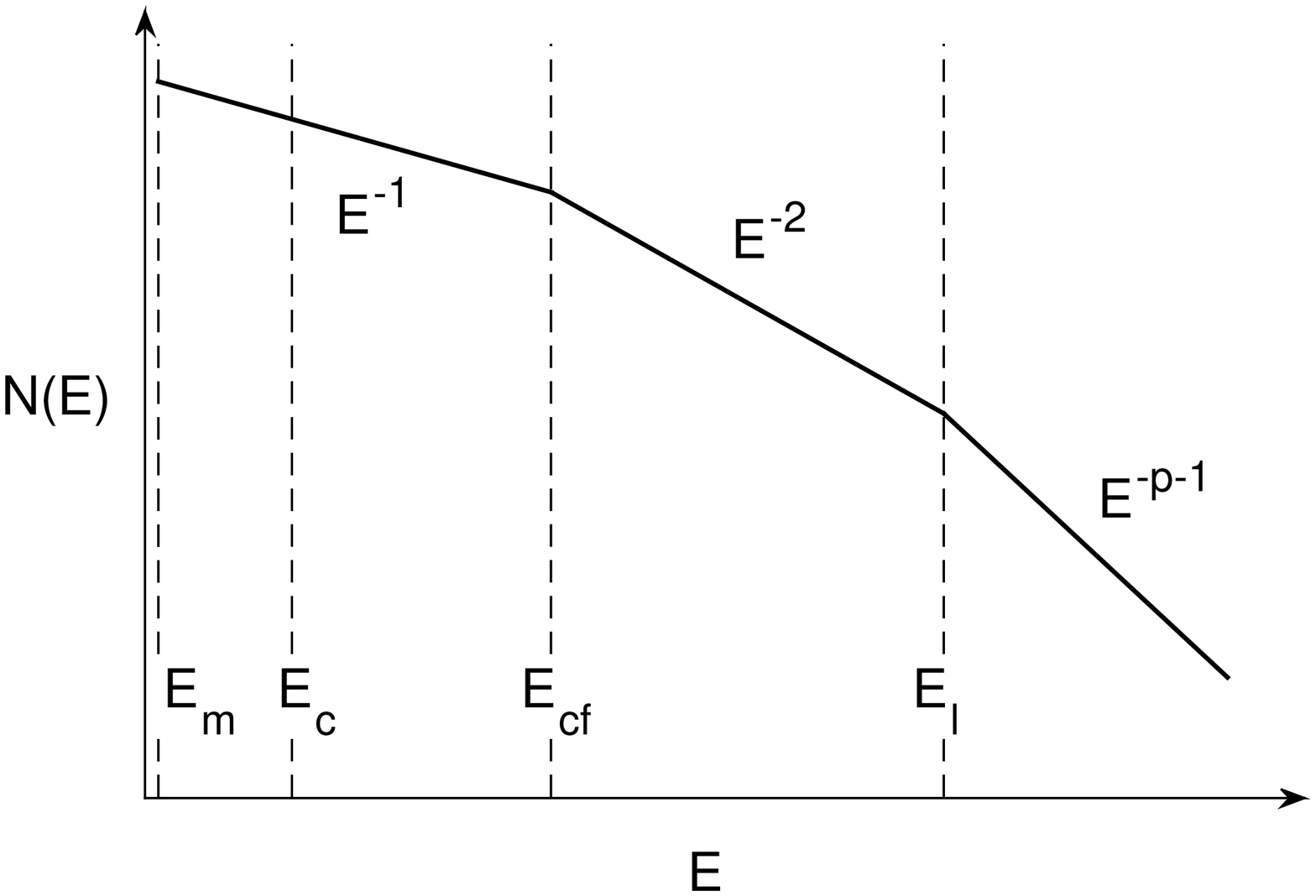}\label{fig: fac2}}
\subfigure[Case (\romannumeral1)]{
   \includegraphics[width=8.5cm]{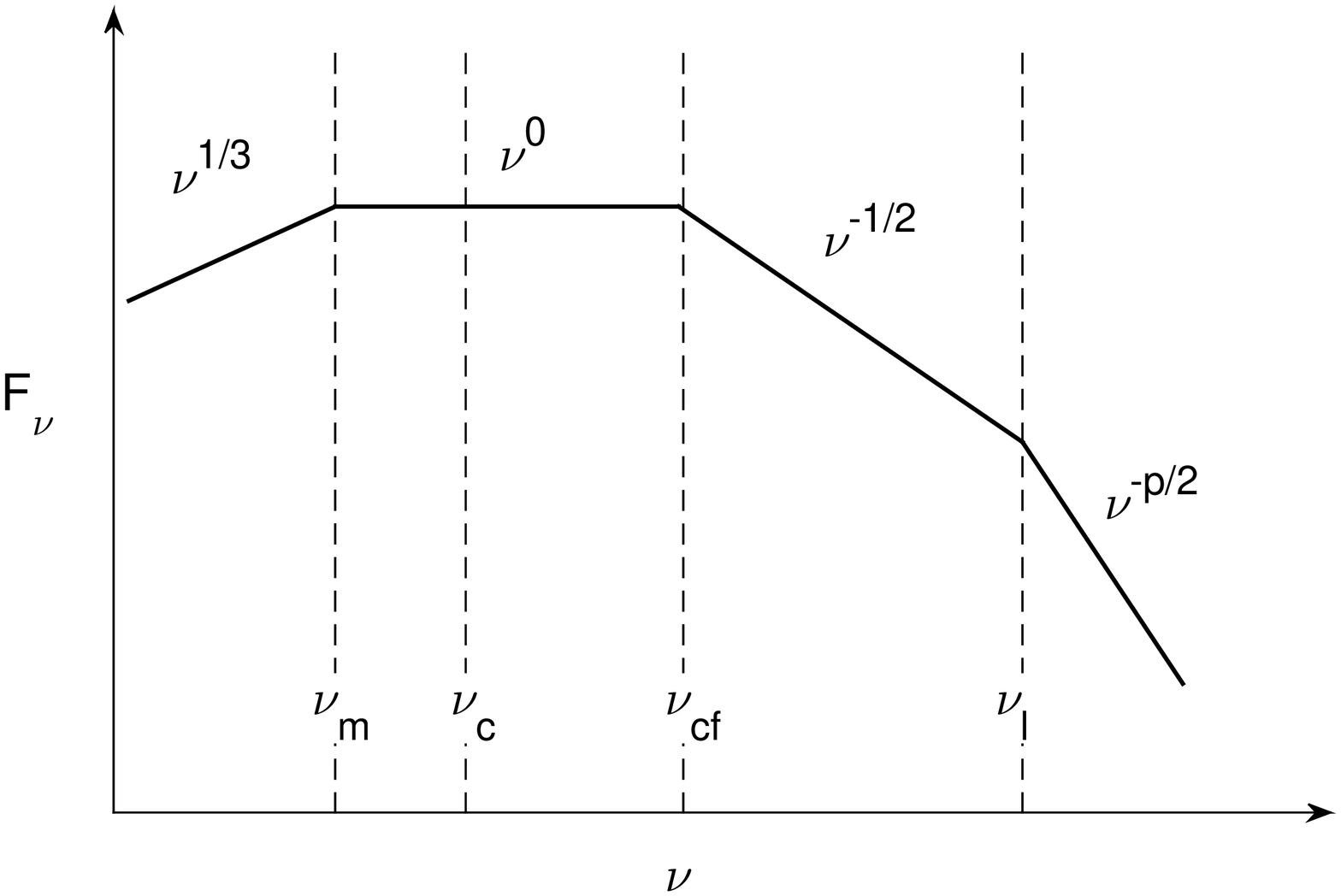}\label{fig: fac2s}}   
   
\subfigure[Case (\romannumeral2)]{
   \includegraphics[width=8.5cm]{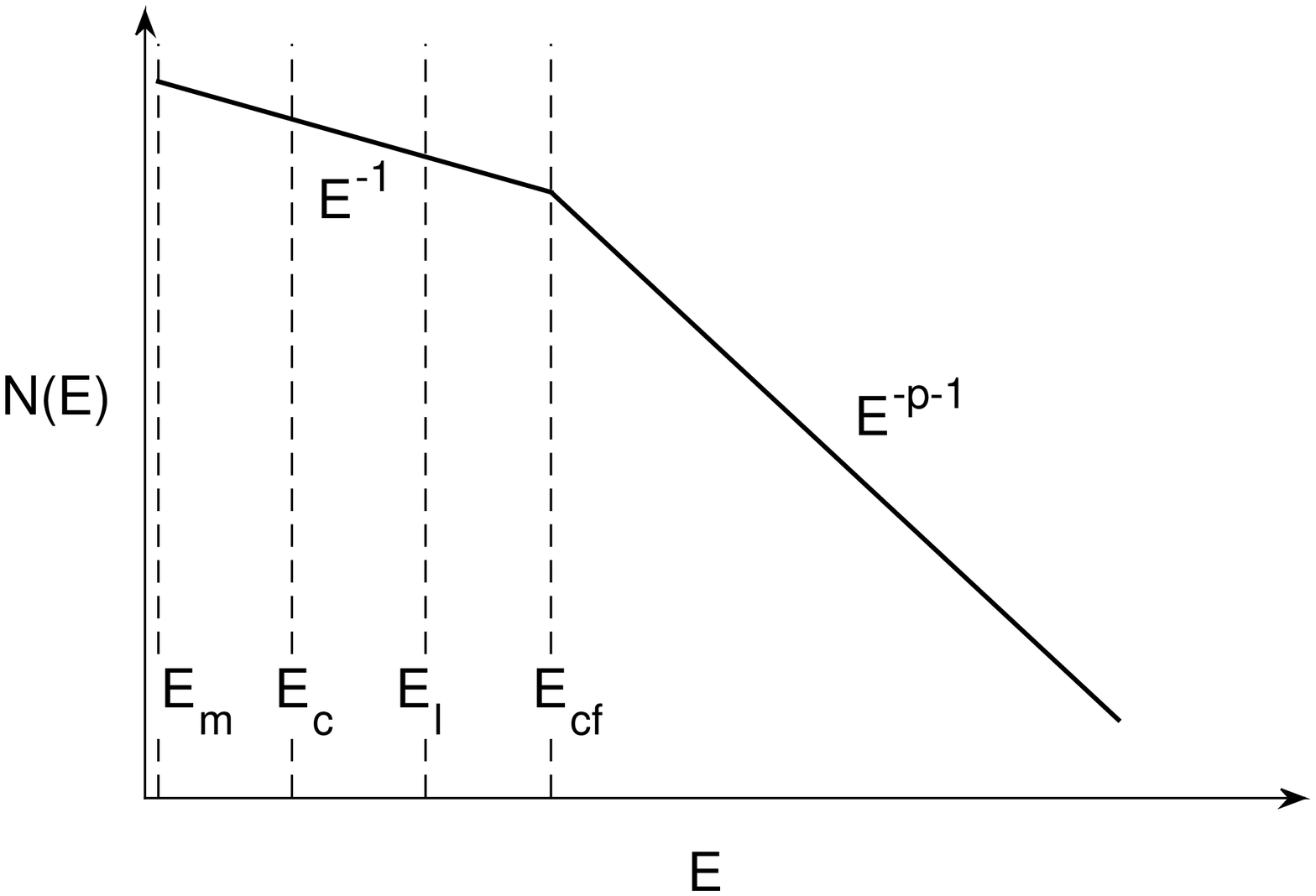}\label{fig: fac1}}
\subfigure[Case (\romannumeral2)]{
   \includegraphics[width=8.5cm]{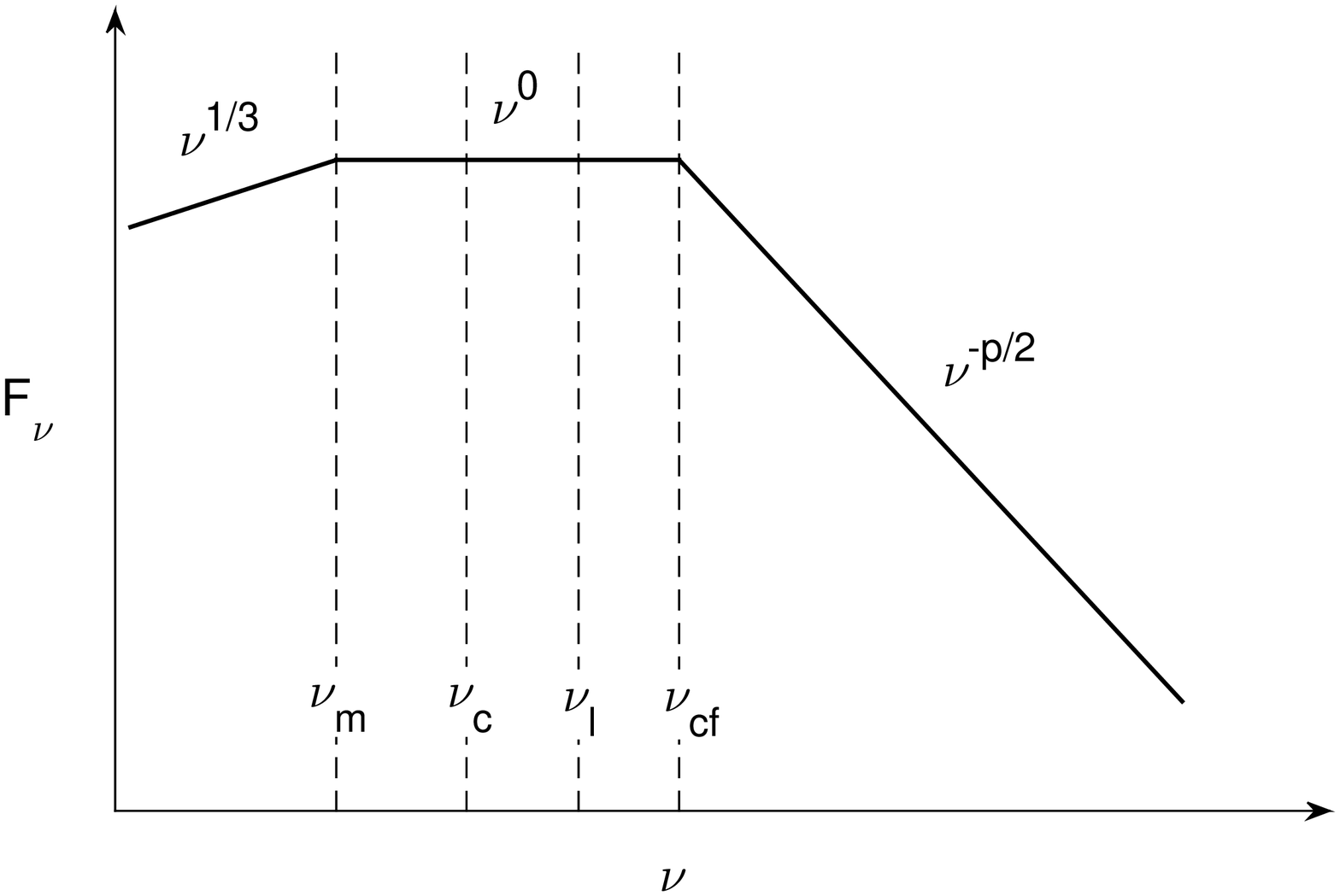}\label{fig: fac1s}}   

\caption{The electron energy spectrum ((a) and (c)) and the corresponding synchrotron spectrum ((b) and (d)) in the fast-cooling regime.  }
\label{fig: fast}
\end{figure*}

In the case with $E_l > E_\text{cf}$ (Case (\romannumeral1), see Fig. \ref{fig: fac2}),
all the injected electrons are in the fast-cooling regime and their spectrum above $E_l$ is given by 
Eq. \eqref{eq: cols}. 
Within the energy range $(E_\text{cf}, E_l)$, the kinetic equation becomes 
\begin{equation}
     \frac{\partial N}{\partial t} =  \beta \frac{\partial (E^2N)}{\partial E} .
\end{equation}
With the boundary condition satisfying a constant particle flux from high to low energies 
driven by the synchrotron cooling
\citep{Ta71,Par95},
the solution at the balance between the particle injection and synchrotron losses is   
\citep{Mel69},
\begin{equation}
    N(E) = N(E_0,0) \frac{E_0^2}{E^2},
\end{equation}
where $E_0$ is the initial electron energy at $t=0$.
{As a result, the minimum energy $E_m$ shifts from $E_l$ to lower energies.
After $E_m$ reaches $E_\text{cf}$}, 
the electron distribution spreads to 
and flattens at
further lower energies due to the ASA.

In the opposite situation with $E_l < E_\text{cf}$ (Case (\romannumeral2), see Fig. \ref{fig: fac1}), 
the electrons at energies beyond $E_\text{cf}$ are subject to the fast cooling, 
whereas those below $E_\text{cf}$ follow a hard spectrum under the effect of the ASA.

In both Case (\romannumeral1) and Case (\romannumeral2), 
although initially the minimum energy of electrons $E_m$ is larger than $E_c$, 
by comparing their time dependences (Eq. \eqref{eq: enecol} and \eqref{eq: eml}),
we see that after reaching below $E_\text{cf}$, $E_m$
can soon become smaller than $E_c$.

The corresponding flux in the two cases 
takes the form as follows (see Fig. \ref{fig: fac2s} and \ref{fig: fac1s}): 
\\
Case (\romannumeral1) $E_c < E_\text{cf} < E_l$,
\begin{subnumcases}
 {F_\nu=}
F_{\nu,\text{max}} \Big(\frac{\nu}{\nu_m}\Big)^\frac{1}{3}  ,~~~~~~~~~~~~~~~~~~~~~~~~\nu<\nu_m,\\
F_{\nu,\text{max}},~~~~~~~~~~~~~~~~~~~~~~~~~~~~\nu_m < \nu <\nu_\text{cf}, \\
F_{\nu,\text{max}} \Big(\frac{\nu}{\nu_\text{cf}}\Big)^{-\frac{1}{2}} ,~~~~~~~~~~~~~~\nu_\text{cf} < \nu < \nu_l, \\
F_{\nu,\text{max}} \Big(\frac{\nu_l}{\nu_\text{cf}}\Big)^{-\frac{1}{2}} \Big(\frac{\nu}{\nu_l}\Big)^{-\frac{p}{2}} ,~~~~~~~~~~\nu_l < \nu.
\end{subnumcases}
Case (\romannumeral2) $E_c < E_l < E_\text{cf}$,
\begin{subnumcases}
 {F_\nu=\label{eq: fcc1}}
F_{\nu,\text{max}} \Big(\frac{\nu}{\nu_m}\Big)^\frac{1}{3}  ,~~~~~~~~~~~~~~~~~~~~~~~~\nu<\nu_m,\\
F_{\nu,\text{max}},~~~~~~~~~~~~~~~~~~~~~~~~~~~~\nu_m < \nu <\nu_\text{cf}, \\
F_{\nu,\text{max}} \Big(\frac{\nu}{\nu_\text{cf}}\Big)^{-\frac{p}{2}} ,~~~~~~~~~~~~~~~~~~~~~~~\nu_\text{cf} < \nu.
\end{subnumcases}
For the $\nu F_\nu$ spectrum, if $p>2$, 
it peaks at $\nu_l$ in Case (\romannumeral1)
and $\nu_\text{cf}$ in Case (\romannumeral2), 
depending on the energy range of the injected electrons.

\subsection{Characteristic spectral parameters}
\label{ssec: csp}

The time-averaged GRB spectra are typically modeled by the empirical Band spectrum 
\citep{Band93},
which is characterized by a low-energy spectral index $\alpha_s$, a break energy $E_b$, and a high-energy spectral index $\beta_s$. 
As shown by \citet{Pre00} based on a large burst sample, 
the indices $\alpha_s$ and $\beta_s$ 
have their distributions centered on $-1$ and $-2.25$, respectively.
The break energy at the transition 
is on the order of $100$ keV,
which corresponds to the peak energy $E_p$ in $\nu F_\nu$ spectrum if $\beta_s < -2 < \alpha_s$.

For the prompt phase of a GRB in the fast-cooling regime, by comparing the above analysis
with the observed spectral characteristics, we 
find that Case (\romannumeral2) in the fast-cooling regime
can satisfactorily reproduce the spectral behavior revealed by observations. 
The hard low-energy spectrum $N(\nu)$ with $\alpha_s \sim -1$ naturally arises over the range $(\nu_m, \nu_\text{cf})$
out of the ASA.
The slope of the injected energy spectrum inferred from $\beta_s = -2.25$ is $p = 2.5$. 
This is close to the typical spectral index of the accelerated electrons in 
relativistic shocks ($p = 2.2$), 
as well as that resulting from the reconnection acceleration ($p = 2.5$)
\citep{DeG05}.

The peak energy $E_p$ (i.e. $E_b$) of the synchrotron spectrum is determined by the cutoff energy $E_\text{cf}$ of the electron energy spectrum. 
Its evaluation has been discussed in Paper \uppercase\expandafter{\romannumeral1}, 
which is again presented here for the sake of completeness. 
By combining Eq. \eqref{eq: a2}, \eqref{eq: betp}, and \eqref{eq: coeg}, $E_\text{cf}$ has the expression as 
\begin{equation}\label{eq: ecfac}
   E_\text{cf} =  \frac{\xi u_\text{tur}}{l_\text{tur} \beta}  = \frac{6\pi \xi (m_e c^2)^2}{\sigma_T B^2 l_\text{tur}}, 
\end{equation}
where $u_\text{tur} \approx c$ for the relativistic turbulence.
Furthermore, we consider that turbulence is mainly driven by the magnetic reconnection,  
with the turbulent energy converted from the magnetic energy released during the magnetic reconnection. 
The turbulence becomes trans-Alfv\'{e}nic when the growing turbulent energy reaches equipartition with the magnetic energy.
According to the evolutionary behavior of the resulting turbulence
\citep{Kow17}
and the expected global decay of magnetic field strength in the emission region due to adiabatic expansion of the jet
\citep{PeZ06,UZ14,ZhL14}, 
we assume that they are anticorrelated by  
\begin{equation}\label{eq: ltb}
   l_\text{tur} = l_\text{0} \Big(\frac{B}{B_0}\Big)^{-\zeta}, ~ \zeta>0,
\end{equation}
with normalization parameters $l_0$ and $B_0$. 
Using the above relation, we can further write Eq. \eqref{eq: ecfac} as, 
\begin{equation}\label{eq: ecfpr}
    E_\text{cf} = \frac{6\pi \xi (m_e c^2)^2}{\sigma_T l_\text{0} B_0^\zeta } B^{\zeta-2}.
\end{equation}
The corresponding photon energy in the observer frame is 
\begin{equation}
\begin{aligned}
    E_{s,\text{cf,obs}}  & = (h \nu_\text{cf})_\text{obs} \\
   &= \hbar \frac{e B}{m_e c} \gamma_{e,\text{cf}}^2 \Gamma (1+z)^{-1}    \\    
   &=  2^{\zeta-\frac{3}{2}}\hbar m_e c^{-\zeta+\frac{9}{2}} e  \bigg(\frac{6\pi \xi}{\sigma_T  l_\text{0} B_0^\zeta}\bigg)^2  \\
   &~~~~~~  \Gamma^{-2\zeta+4} (1+z)^{-1} L^{\zeta-\frac{3}{2}} r^{-2\zeta+3},
\end{aligned}
\end{equation}
with $\gamma_{e,\text{cf}} = E_\text{cf}/(m_e c^2)$, 
the Planck constant $h$, the electron charge $e$, the redshift $z$,
and the substitution
\citep{Zhang02}, 
\begin{equation}\label{eq: zmag}
    B = \sqrt{\frac{2}{c}} L^\frac{1}{2} r^{-1} \Gamma^{-1}.
\end{equation}
By adopting $\Gamma=100 \Gamma_{2}$ for the bulk Lorentz factor,
$L=10^{52} {\rm erg~s^{-1}} L_{52}$ for the total isotropic luminosity,
$r=10^{15} \text{cm} r_{15}$ for the radius of the emission region, 
$\zeta \simeq 2.1$ based on the observational constraints  
\citep{Lia15}, 
as well as $\xi =10^4$, $B_0 = 10^5$G, $l_0 = 2\times10^9$cm, 
we can then evaluate $E_{s,\text{cf,obs}}$, 
\begin{equation}
   E_{s,\text{cf,obs}} \simeq 385 \text{keV}~ \Big(\frac{1+z}{2}\Big)^{-1} \Gamma_2^{-0.2} L_{52}^{0.6} r_{15}^{-1.2},
\end{equation}
which is compatible with the observationally revealed peak energy.

One necessary condition for Case (\romannumeral2) in the fast-cooling regime
is $E_c < E_\text{cf}$, which is equivalent to (Eq. \eqref{eq: coeg}, \eqref{eq: enecol})
\begin{equation}\label{eq: tae}
   t > \frac{1}{a_2} = \frac{l_\text{tur}}{\xi u_\text{tur}}.             
\end{equation}
By using Eq. \eqref{eq: ltb} and \eqref{eq: zmag}, we find 
\begin{equation}\label{eq: ta2}
\begin{aligned}
    t & > \frac{l_0 B_0^\zeta}{\xi c} \Big(\frac{2}{c}\Big)^{-\frac{\zeta}{2}} L^{-\frac{\zeta}{2}} r^\zeta \Gamma^\zeta   \\
      & =1.3\times10^{-3} \text{s}~ L_{52}^{-1.05} r_{15}^{2.1} \Gamma_2^{2.1},
\end{aligned}
\end{equation}
as the time required for $E_c$ to decrease below $E_\text{cf}$. 
In the observer frame, it becomes 
\begin{equation}\label{eq: tesof}
    t_\text{obs} > 1.3\times10^{-5} \text{s}~ L_{52}^{-1.05} r_{15}^{2.1} \Gamma_2^{1.1},
\end{equation}
which is so short that we can safely assume that the relation $E_c < E_\text{cf}$ always holds.

The other condition $E_l < E_\text{cf}$ implies that 
the energy range of the injected electrons should be broad enough to reach down to $E_\text{cf}$. 
Given the estimate of $E_\text{cf}$, there is (Eq. \eqref{eq: ecfpr}, \eqref{eq: zmag}) 
\begin{equation}\label{eq: conlcf}
\begin{aligned}
     E_l < E_\text{cf} &= \frac{6\pi \xi (m_e c^2)^2}{\sigma_T l_\text{0} B_0^\zeta } 
                                  \Big(\frac{2}{c}\Big)^\frac{\zeta-2}{2} L^\frac{\zeta-2}{2} r^{2-\zeta} \Gamma^{2-\zeta} \\
                                &= 4.6 \text{GeV} L_{52}^{0.05} r_{15}^{-0.1} \Gamma_2^{-0.1}.
\end{aligned}     
\end{equation}
Starting from $E_l$, the minimum energy of electrons further decreases with time driven by the ASA. 
Following Eq. \eqref{eq: eml}, we can write 
\begin{equation}
\begin{aligned}
    E_m &= E_l \exp (-2 \sqrt{a_2 t}) \\
            &= E_l \exp \Big[-2 \Big(\frac{\xi c}{l_0 B_0^\zeta}\Big)^\frac{1}{2} \Big(\frac{2}{c}\Big)^\frac{\zeta}{4} L^\frac{\zeta}{4} r^{-\frac{\zeta}{2}}
                  \Gamma^{\frac{-\zeta+1}{2}} t_\text{obs}^\frac{1}{2} \Big]        \\
            &= E_l \exp \Big[-558  L_{52}^{0.525} r_{15}^{-1.05} \Gamma_2^{-0.55} t_{\text{obs},0}^{0.5}\Big]     ,
\end{aligned}
\end{equation}
where $t_\text{obs} = 1 \text{s} t_{\text{obs},0}$, and the expression of $a_2$ can be extracted from Eq. \eqref{eq: tae} and \eqref{eq: ta2}. 
We see that $E_m$ rapidly decreases 
to a negligibly small value compared with $E_l$, forming an extended hard energy spectrum of electrons. 
It leads to the photon energy: 
\begin{equation}
\begin{aligned}
  &  E_{s,m,\text{obs}} = \hbar \frac{e B}{m_e c} \gamma_{e,m}^2 \Gamma (1+z)^{-1}    \\ 
                               &= E_{s,l,\text{obs}}\exp \Big[-1.1\times10^3  L_{52}^{0.525} r_{15}^{-1.05} \Gamma_2^{-0.55} t_{\text{obs},0}^{0.5}\Big]   
\end{aligned}
\end{equation}
where 
\begin{equation}
   E_{s,l,\text{obs}} = \hbar \frac{e B}{m_e c} \gamma_{e,l}^2 \Gamma (1+z)^{-1}  , 
\end{equation}
with $\gamma_{e,m} = E_m / (m_e c^2)$ and $\gamma_{e,l} = E_l / (m_e c^2)$.
The time evolution of $E_m$ discussed above is 
solely affected by the ASA. 
If there is additional systematic acceleration also coming into play in the emission region, its behavior 
can be found in Appendix \ref{app}.

\section{Discussion}

In the standard model for synchrotron emission from shock-accelerated electrons, 
the sites for instantaneous acceleration and radiation are separate, 
under the assumptions of a highly efficient acceleration and thus a negligible cooling effect at relativistic shocks. 
In a more realistic situation, 
the instantaneously accelerated electrons, after being injected into the magnetized and turbulent medium,
are likely to 
experience further continuous reacceleration and radiation simultaneously 
\citep{Llo00,Asano09,Asano15}.
In the model discussed in Paper \uppercase\expandafter{\romannumeral1} and this paper, 
we introduce the ASA for the electrons injected into the trans-Alfv\'{e}nic turbulence.
This diffusive acceleration acts against the synchrotron cooling 
and generates a hard spectrum up to the cutoff energy 
with balanced acceleration and cooling rates (Eq. \eqref{eq: coeg}). 
The stochastic nature of the ASA originates from the equipartition between the turbulent kinetic and magnetic energies
in the trans-Alfv\'{e}nic turbulence,
and it does not impose the resonance condition on particles to be accelerated. 
In contrast, 
the conventionally adopted 
stochastic acceleration through the pitch-angle scattering of particles by magnetic fluctuations is 
disregarded here, 
because the resonant condition may not be satisfied with a small Larmor radius of particles at a strong magnetization. 
Moreover, even in the presence of sufficiently small-scale magnetic turbulence, the resonant scattering is 
very inefficient due to the prominent turbulence anisotropy especially at small scales
\citep{YL02}.
At high energies away from the cutoff energy, without significant ASA, 
a steep spectrum of the injected electrons arises under the fast-cooling effect.

As discussed above, 
the underlying physical process of the particle injection is the instantaneous acceleration in a separate region 
different from the ASA region. 
The possible injection mechanisms include the shock acceleration as adopted in the classical GRB model 
\citep{Ree94}, 
which has the acceleration timescale as the particle gyroperiod 
\citep{Byk96,Bed00}.
They could also be the reconnection acceleration 
(e.g. \citealt{DeG05}), 
or the hadronic injection via $pp$ and $p\gamma$ reactions
(e.g., \citealt{Mur12}). 
The instantaneous acceleration should be able to produce a power-law spectrum of electrons over a relatively broad energy range 
with the lower and upper energy bounds 
below and above $E_\text{cf}$, respectively (see Fig. \ref{fig: fac1}).
A detailed model for the injection mechanism connecting both the energy dissipation 
and particle acceleration will be constructed in our future work.

Although the ASA can successfully explain the typical value of the low-energy spectral index 
$\alpha=-1$, 
observations suggest that $\alpha$ has a wide distribution around $-1$
\citep{Ghir03}.
\citet{Llo00}
pointed out that the dispersion in $\alpha$
can come from the finite bandwidth of the instrument and the spectral fitting effects.  
It is noticeable that $\alpha = -1$ in our analysis is given in the asymptotic limit below the cutoff frequency $\nu_\text{cf}$. 
Above and near $\nu_\text{cf}$, the spectrum can be piled up 
due to the interplay between acceleration and cooling
\citep{Sch84,Sch85}. 
The spectra harder than $\alpha = -1$ 
can be also attributed to the synchrotron spectrum with $\alpha = -2/3$
produced by monoenergetic electrons. 
Beyond this ``line of death" of the synchrotron emission, 
different physical models based on e.g.,  synchrotron self-absorption
\citep{Pree98}, 
small pitch-angle synchrotron emission 
\citep{Llo00,Lio02}, 
or inverse Compton scattering 
\citep{Lia97}
have been developed for understanding the spectral hardness.

A caveat in our analysis is that the decay of $B$ was not incorporated in the kinetic equation of electrons, 
but only added later in our evaluation of the characteristic spectral parameters (Section \ref{ssec: csp}). 
In a more realistic scenario, 
\citet{UZ14}
(see also \citealt{Geng17})
showed that 
a decaying $B$ can affect the fast-cooling process and results in a hard energy spectrum of the cooled electrons
with the index $\sim -1$ instead of $-2$. 
Under this consideration, 
a hard electron spectrum would arise within the energy range $(E_\text{cf}, E_l)$ in Case (\romannumeral1)
in the fast-cooling regime, 
which provides an alternative explanation for the Band function without invoking the ASA at lower energies. 
Notice that the ASA generated electron spectrum would also be modified 
if the temporal dependence of $B$ is taken into account in Eq. \eqref{eq: soasi}. 
These effects will be addressed in future work.

\section{Summary}

As an attempt to fully understand the origin of GRB Band function within the framework of synchrotron
radiation mechanism, 
we have analyzed the time evolution of the injected electron spectrum under the effects of both the ASA and the synchrotron cooling.   
Their relative importance depends on the energy range of interest.

The cutoff energy $E_\text{cf}$ corresponds to the balance between the ASA and the synchrotron cooling. 
At energies below $E_\text{cf}$, 
the ASA plays a dominant role and 
ensures a rapid development of a flat energy distribution of electrons, 
in spite of the steady injection of a steep electron spectrum. 

At higher energies,
the synchrotron cooling effect is manifested. The injected high-energy electrons in the slow- 
and fast-cooling regimes evolve to various spectral forms, 
which are all steeper than the low-energy spectral component.

Among different scenarios in the fast-cooling regime, if the lower energy limit of the injected electrons is below $E_\text{cf}$
(Eq. \eqref{eq: conlcf}), 
the ASA can efficiently harden the injected distribution below $E_\text{cf}$ and spread it out 
to further lower energies.
The resulting low- and high-energy spectral components can together reproduce the Band spectrum of GRBs,
with $E_\text{cf}$ corresponding to the break energy.

\section*{Acknowledgement}

S.X. acknowledges the support for Program number HST-HF2-51400.001-A provided by NASA through a grant from the Space Telescope Science Institute, which is operated by the Association of Universities for Research in Astronomy, Incorporated, under NASA contract NAS5-26555.
This work is partially supported by Project funded by the Initiative Postdocs Supporting Program (No. BX201600003) and the China Postdoctoral Science Foundation (No. 2016M600851). 
Y.-P.Y. is supported by a KIAA-CAS Fellowship.

\appendix

\section{The electron energy spectrum affected by both the stochastic and systematic acceleration }
\label{app}

The kinetic equation including both the stochastic and systematic acceleration processes is 
\begin{equation}\label{eq: kincon}
\begin{aligned}
   \frac{\partial N}{\partial t} =&  a_2 \frac{\partial}{\partial E} \Big(E\frac{\partial (EN)}{\partial E}\Big)
   - a_{1} \frac{\partial (EN)}{\partial E}  ,
\end{aligned}
\end{equation}
where $a_1$ accounts for the systematic Fermi acceleration. 
By again substituting the variables $f=EN$, $x=\ln E$, $\tau=a_2 t$, 
the above equation can be rewritten as 
\begin{equation}\label{eq: rewt}
     \frac{\partial f}{\partial \tau} = \frac{\partial ^2 f}{\partial x^2} - a_{12} \frac{\partial f}{\partial x}, 
\end{equation}
where $a_{12} = a_1/a_2$. 
Following the similar algebra as in 
\citet{Mel69}, 
by introducing 
\begin{equation}
    X = x - a_{12} \tau,
\end{equation}
 we find 
\begin{equation}
       \frac{\partial f}{\partial \tau}  = \frac{\partial f(X,\tau)}{\partial \tau}  - a_{12} \frac{\partial f(X,\tau)}{\partial X}, 
\end{equation}
and
\begin{equation}
    \frac{\partial f}{\partial x} = \frac{\partial f(X,\tau)} {\partial X}, ~~ \frac{\partial ^2 f}{\partial x^2} = \frac{\partial ^2 f(X,\tau)}{\partial X^2} .
\end{equation}
Then Eq. \eqref{eq: rewt} can be further simplified,
\begin{equation}
    \frac{\partial f}{\partial \tau} = \frac{\partial^2 f}{\partial X^2}. 
\end{equation}
The solution to this diffusion equation is 
(e.g., \citealt{Ev98}),
\begin{equation}
     f(X,\tau) =  \frac{1}{2 \sqrt{\pi \tau}}  \int_{y_l}^{y_u}   \exp\Big[-\frac{(X - y)^2}{4 \tau}\Big] f(y, 0) dy .
\end{equation}
The initial condition is given by $f(y,0)$ within the finite range ($y_l$, $y_u$), where $y_l = \ln E_l$ and $y_u = \ln E_u$.
Therefore we obtain the energy spectrum as 
(see e.g. \citealt{Kar62,Mel69}),
\begin{equation}\label{eq: esba}
    N(E,\tau)  = E^{-1}  \frac{1}{2 \sqrt{\pi \tau}}  \int_{y_l}^{y_u}   \exp\Big[-\frac{(\ln E - a_{12} \tau - y)^2}{4 \tau}\Big] f(y, 0) dy .
\end{equation}
When $a_{12}=0$, i.e., $a_1=0$, it recovers the result with only the stochastic acceleration being considered (see Eq. \eqref{eq: p1ns}).
When $a_{12}>0$, given the time dependence of the peak position of the gaussian function in Eq. \eqref{eq: esba}, 
we see that electrons initially at $E_0$ spread out in the energy space and can reach $E_\tau$ after a time $\tau$, 
\begin{equation}\label{eq: enevl}
    E_\tau = E_0 \exp (\pm 2 \sqrt{\tau} + a_{12} \tau) .
\end{equation}
By taking logarithms of both sides, it can be expressed in a more physically transparent way,
\begin{equation}
    \ln E_\tau = \ln  E_0 \pm 2 \sqrt{\tau} + a_{12} \tau.
\end{equation}
Different from the case of purely stochastic acceleration, which leads to a random walk in the $\ln E$ space, 
the combination of both the stochastic and systematic acceleration drives an asymmetric diffusion about $\ln E_0$. 
With a linear dependence on time,  
the second term in the exponential function in Eq. \eqref{eq: enevl}
becomes dominant over the first one at a large $\tau$, irrespective of the value of $a_{12}$ ($a_{12}>0$). 
More specifically, when $\tau > 4/a_{12}^2$, there is always $E_\tau > E_0$, 
and the lower limit of the energy $E_m$ increases with time.

The asymptotic behavior of $N(E,\tau)$ at a sufficiently large $\tau$ is
\begin{equation}
    N(E,\tau) \approx  E^{-1}  \frac{1}{2 \sqrt{\pi \tau}} \exp\Big(-\frac{a_{12}^2}{4}\tau\Big)    \int_{y_l}^{y_u}    f(y, 0) dy .
\end{equation}
Compared with the spectrum governed only by the stochastic acceleration, 
under the combined effect of the two acceleration mechanisms, 
the spectrum is still flat, but  
the number of electrons exponentially decreases as they shift to higher energies. 

When the synchrotron cooling is taken into account, 
around the energy with comparable acceleration and cooling rates, 
it acts in concert with the systematic acceleration in 
narrowing down the energy range and piling up the electron distribution.

\bibliographystyle{apj.bst}
\bibliography{xu}

\end{document}